\documentclass[apl,preprint]{revtex4-1}

\usepackage{graphicx}
\usepackage{amsfonts}
\usepackage{floatrow}

\usepackage{amssymb,amsmath}

\setcitestyle{super}

\draft 
\begin{document}

\title{Position and Mode Dependent Coupling of Terahertz Quantum Cascade Laser Fields to an Integrated Diode} 

\newcommand{\tc}{\textcolor[rgb]{1,0,0}}
\newcommand{\dIdV}{${\partial^2 I_D} / {\partial V_{D}^2}$\,}
\newcommand{\dIdT}{${\partial I_D} / {\partial T}$\,}
\newcommand{\diQCL}{$\delta I_{QCL}$}
\newcommand{\dIddIQCL}{$\delta I_{D} / \delta I_{QCL} \:$}
\author{Gregory C. Dyer}
\affiliation{Sandia National Laboratories, P.O. Box 5800, Albuquerque, NM 87185 USA}
\author{Christopher D. Nordquist}
\author{Michael J. Cich}
\altaffiliation{Now at Soraa, Freemont, California, 94555 USA}
\affiliation{Sandia National Laboratories, P.O. Box 5800, Albuquerque, NM 87185 USA}
\author{Troy Ribaudo}
\affiliation{Sandia National Laboratories, P.O. Box 5800, Albuquerque, NM 87185 USA}
\author{Albert D. Grine}
\affiliation{Sandia National Laboratories, P.O. Box 5800, Albuquerque, NM 87185 USA}
\author{Charles T. Fuller}
\affiliation{Sandia National Laboratories, P.O. Box 5800, Albuquerque, NM 87185 USA}
\author{John L. Reno}
\affiliation{Sandia National Laboratories, P.O. Box 5800, Albuquerque, NM 87185 USA}
\author{Michael C. Wanke}
\email[]{mcwanke@sandia.gov}
\affiliation{Sandia National Laboratories, P.O. Box 5800, Albuquerque, NM 87185 USA}
\date{\today}
\begin{abstract}
A Schottky diode integrated into a terahertz quantum cascade laser waveguide couples directly to the internal laser fields. In a multimode laser, the diode response is correlated with both the instantaneous power and the coupling strength to the diode of each lasing mode. Measurements of the rectified response of diodes integrated in two quantum cascade laser cavities at different locations indicate that the relative diode position strongly influences the laser-diode coupling.
\end{abstract}
\maketitle 
Quantum cascade lasers (QCLs) may be considered one of the most remarkable achievements in quantum engineering due to both the intensity and the broad tailorability of their emission\cite{Faist1994}. Since the operating range of these unipolar, intersubband lasers was extended to the terahertz (THz) band of the spectrum\cite{Kohler2002}, a variety of applications requiring a compact high-power ($>$mW) source between 1-5 THz have become accessible\cite{Lee2007,LeeAWM2006,Kim2006,Hubers2006,Dean2008,Behnken2008}. Of particular interest is the use of a THz QCL as a local oscillator (LO) for heterodyne mixing\cite{Barbieri2004,Gao2005,Hubers2005,Lee2008,Richter2008}. THz QCLs provide ample power for mixing; however, it is non-trivial to efficiently couple the THz LO power from a QCL to a mixer such as a planar Schottky diode. One possible solution is to directly integrate a Schottky diode mixer into the core of a THz QCL to create a THz transceiver\cite{Wanke2010}.

We previously observed the direct coupling of the internal QCL fields to an integrated diode\cite{Dyer2013OPEX}, however, several questions concerning the precise nature of this coupling remain open. For practical applications, the response of a Schottky diode mixer should be linear in both the LO and signal field amplitudes. However, prior measurements suggested that both the mode structure and the instantaneous power of the laser may affect the laser-diode coupling\cite{Dyer2013OPEX} and lead to a non-linear response to the QCL (LO) power. In this letter we examine how the rectified response of Schottky diodes embedded into the core of THz QCLs depends upon diode position and QCL bias current. To determine the effect of diode position upon the diode's coupling with the laser fields, we compare the rectified response of diodes with different relative positions in the laser waveguide to the emission spectra of two otherwise identical 2.8 THz QCL transceivers. 

The studied THz QCLs have a Schottky diode embedded into the core of the 3 mm long by 170 $\mu$m wide waveguide, as illustrated in Fig. 1. Both transceivers were cleaved from the same row of the processed die, and thus have identical cavity lengths. Sample A has the diode located by design at the center of the QCL waveguide relative to the laser facets, 1.5 mm from both facets. Sample B has the diode shifted +4 $\mu$m from that of the diode in Sample A. Given the slight uncertainty of the cleave planes relative to the diode position, the exact locations of the diodes in Samples A and B may differ from design. But the relative positions of the two diodes are fixed by the device layouts.
 
\begin{figure}
\includegraphics[scale=.45]{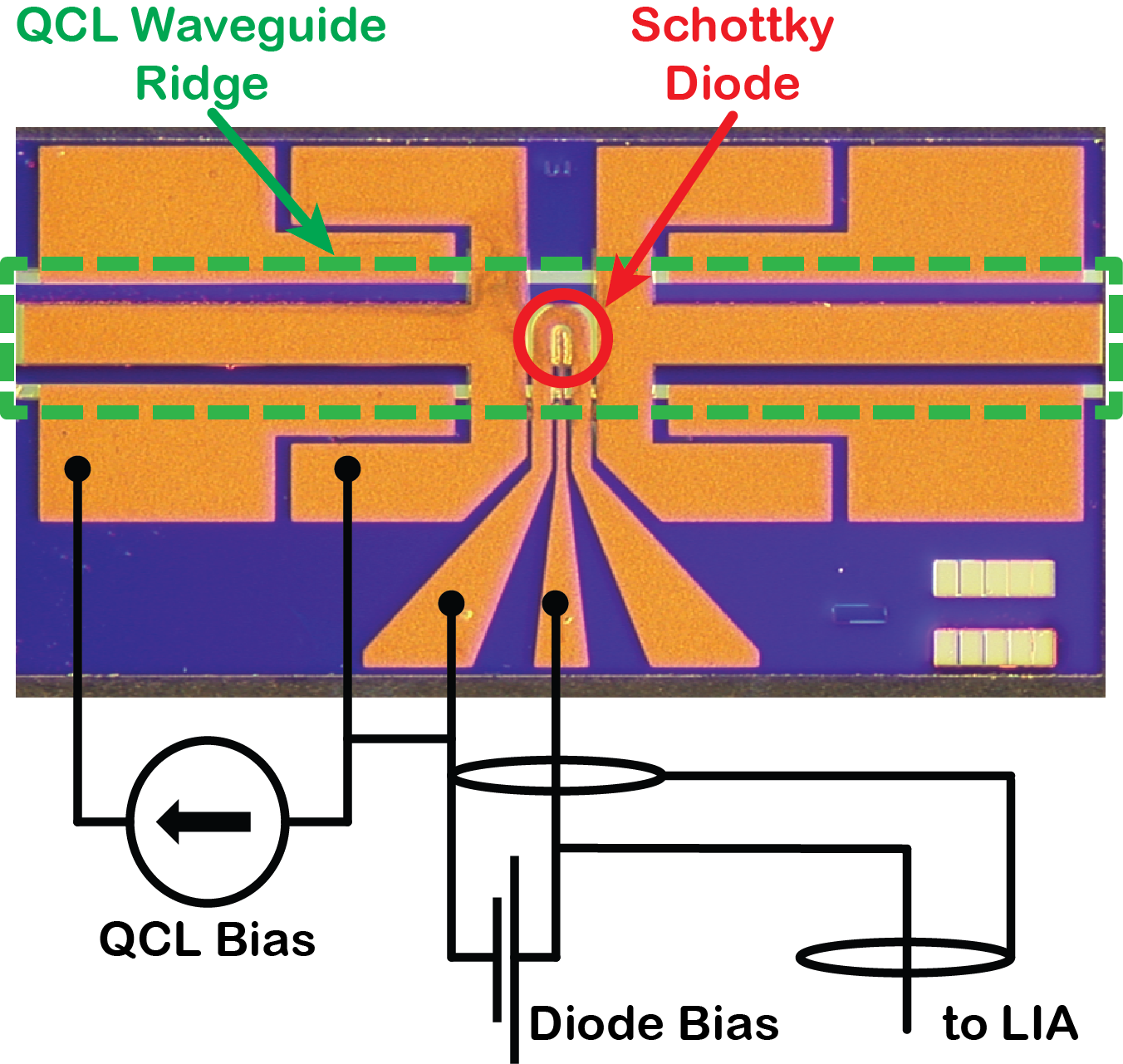}
\caption{The measurement and bias circuits used to characterize the QCL transceivers are shown schematically on a micrograph of the 170 $\mu$m wide by 3 mm long device.}
\end{figure}

\begin{figure}
\includegraphics[scale=0.95]{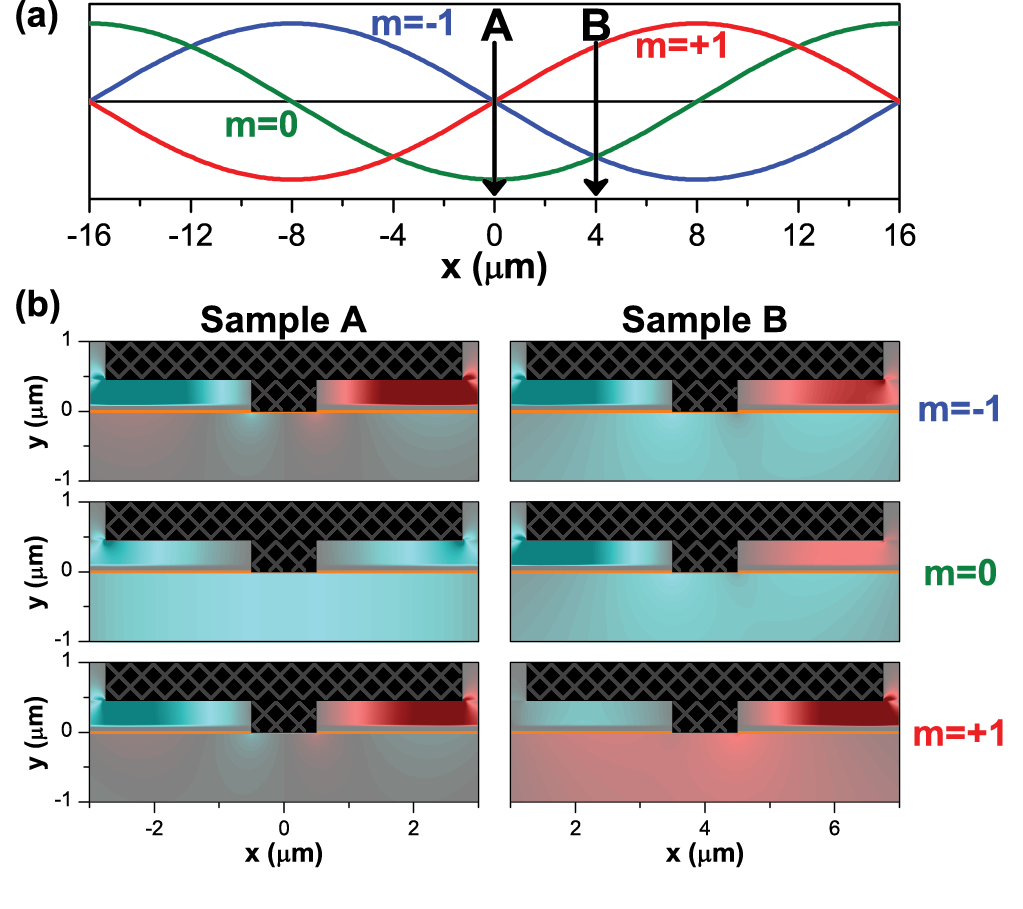}
\caption{(a) Field amplitudes in the central region of the QCL waveguide for FP modes 0 and $\pm$1 with the relative positions of the diodes of Sample A and B indicated. (b) Spatial distributions of $E_y$ near the diode for FP modes 0 and $\pm$1 of Sample A and B are shown.  The diode metalization is indicated by the cross-hatched regions. The line at y=0 indicates the boundary between the doped cap layer of the heterostructure and the QCL active region. Grey areas correspond to zero field.}
\end{figure}


Rectified and intermediate frequency (IF) signals result from the coupling of THz laser fields to a Schottky diode.  If only nearest-neighbor modes in a Fabry-Perot laser (FP) cavity separated by the angular frequency $\omega_{FP}$ are considered, the rectified and IF signals, respectively, can be defined in terms of the nonlinearity of the diode's DC transport,
\begin{equation}
\delta I_{D,rect} = \frac{1}{2} \; \frac{\partial^2 I_{D}}{\partial V_{D}^2} \sum_m \frac{1}{2}  V_m^2
\end{equation}
\begin{equation}
\delta I_{D,IF} = \frac{1}{2} \; \frac{\partial^2 I_{D}}{\partial V_{D}^2} \sum_m V_m V_{m+1}cos(\omega_{FP} t +\phi_m -\phi_{m+1})
\end{equation}
where $V_m$ is the THz voltage generated across the diode by the $m^{th}$ FP mode and $\phi_m$ is the phase of the $m^{th}$ FP mode.  For the transceivers reported here the IF signal (at frequency $\omega_{FP}$) as defined in Eq. 2  is in the microwave $K_u$ band ($\approx$ 13 GHz)\cite{Wanke2010,Wanke2011OPEX}, and is directly measurable using standard RF techniques. However, when more than two modes are present the IF amplitude depends not only upon the field amplitudes coupled to the diode, but also their relative phases. A rectified response, in contrast, depends only upon the instantaneous power of each FP mode coupled to the diode and is therefore easier to analyze than the IF response given the unknown phase relations between terms in Eq. 2. 

The THz QCL transceiver bias and measurement schematic is shown in Fig. 1.  An ILX Lightwave LDX-3232 current supply was used to current bias the QCLs, while a Keithley 238 source-measure unit provided a DC voltage bias to the diode. The diodes were biased to $V_{D}=+0.6$ V and $V_{D}=+0.8$ V in Samples A and B, respectively. Above lasing threshold, the emission from one laser facet is measured using a Bruker FTIR spectrometer. The rectified diode response is measured as follows. Using the oscillator output of a Stanford Research 830 lock-in amplifier (LIA) to control the LDX-3232 current supply, a small AC modulation in the QCL current is applied at 490 Hz.  This in turn modulates the laser gain, internal field of each laser mode, and emission. The voltage across the diode measured at the 490 Hz modulation frequency corresponds to the slope of the diode rectified response, \dIddIQCL\cite{Dyer2013OPEX}. All measurements were performed at 30 K. 

Because the diode is subwavelength--1 $\mu$m diameter in comparison to the $\approx$30 $\mu$m wavelength of the 2.8 THz radiation in the laser cavity--the diode is expected to sample the local rather than spatially averaged laser fields. We therefore expect that different modes will couple differently to the diode depending on the spatial overlap of the mode and diode. Fig. 2(a) shows a cartoon to help explain the reasoning behind this expectation. The figure depicts the standing wave amplitudes of three FP modes of equal intensity in the center portion of a 3 mm long passive cavity. The modes are labeled as 0, and $\pm$1. Using a commercial FDTD package (Lumerical), we also calculated the field distributions near the integrated diode for these three modes in a passive waveguide based upon the fabricated QCL transceiver design. The spatial distributions of the field component $E_y$ are shown in Fig. 2(b) for the relative diode positions in Samples A and B and agree with the qualitative picture in Fig. 2(a).

For a 1 $\mu$m diode at the cavity center as in Sample A, the FDTD calculations show that mode 0 has an antinode at the position of the diode and its fields would be expected to couple strongly to the diode. Modes $\pm$1 however have nodes at the cavity center and thus are predicted to drive the diode less effectively. If the diode is offset by 4 $\mu$m ($\approx \lambda$/8 of the wavelength) as in Sample B, all three modes have nearly the same field amplitude at the position of the diode and therefore should all have similar coupling to the diode.


To quantify the relative power of each FP mode as a function of QCL current, THz emission spectra were measured with an FTIR. The emission spectra of Samples A and B as a function of QCL current are plotted in false color in Figs. 3(a) and (b), respectively. At threshold, only a single laser mode turns on. As the QCL current is increased, additional lasing modes separated by the FP cavity frequency spacing, $\omega_{FP}/2\pi=c/2n_{eff}L_{cavity} \approx 0.013$ THz, become active. The threshold mode of Sample A, $\nu_0=$2.837 THz, is defined as mode 0. FP modes higher in frequency than mode 0 are given positive indices, while modes lower in frequency are given negative indices. Up to modes $\pm$5 are observed for 780 mA QCL bias current applied to Sample A.

Since the lasers are nominally identical (assuming the diode does not perturb the laser), we should expect them to perform similarly. Consistent with this expectation, the frequencies of the FP modes in both lasers are nearly identical, although the threshold mode in Sample B, $\nu_{-1}$=2.823 THz, corresponds to the frequency of the $m$=-1 mode in Sample A. Additional differences between the two transceivers are further illustrated in Figs. 3(c) and (d) where the emitted power P$_m$ of each of the $m$ FP modes is shown as a function of QCL bias current for Samples A and B, respectively. The modes in Sample A appear to pair up. For a given pair $\pm m$, both plus and minus modes have similar behavior, i.e. they have similar thresholds, similar slope efficiencies when they first turn on, and similar amplitudes. In Sample B, the $\pm m$ modes do not pair. They have different threshold currents and their emitted powers differ as a function of QCL current. Based on Eq. 1, these differences in mode amplitudes should be reflected in the rectified diode response.

\begin{figure}
\includegraphics[scale=0.95]{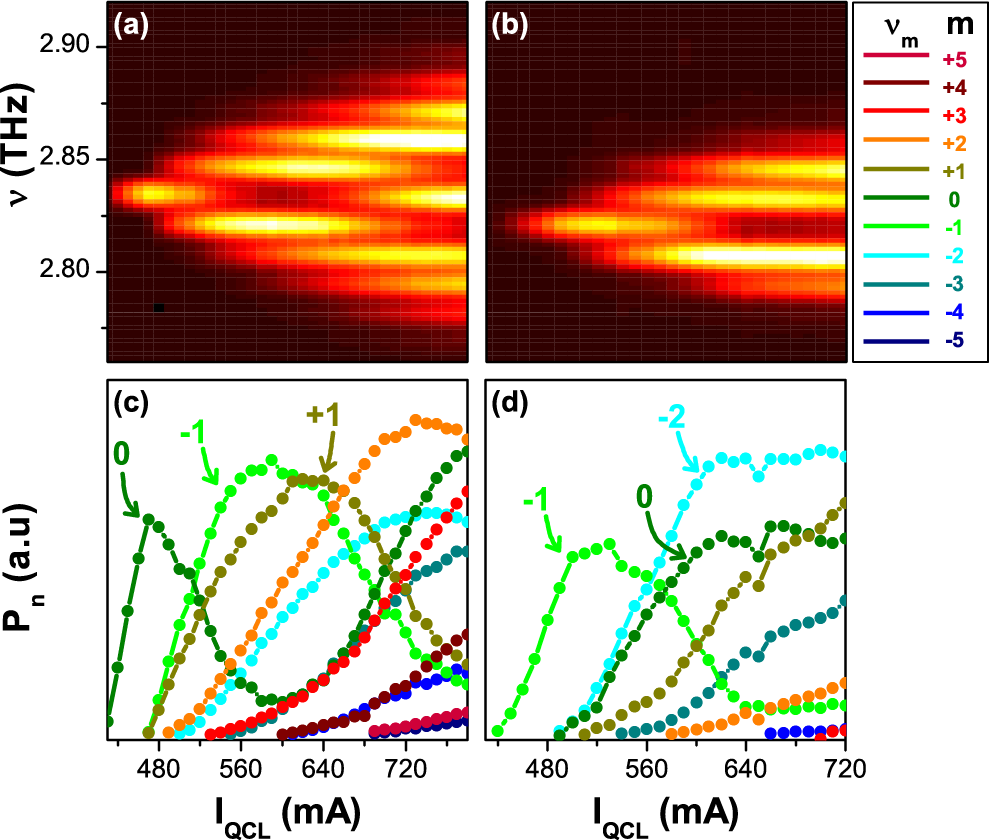}
\caption{The THz emission of Samples A and B is shown in (a) and (b), respectively, with bright areas corresponding to high emission intensity and dark to zero emitted power. We assigned a mode number to each particular frequency with mode 0 corresponding to the frequency of the threshold mode in Sample A. The peak emission intensity of each mode $\pm$m is shown for Samples A and B in (c) and (d). The corresponding color coding to indicate the frequency of each mode $\pm$m is shown to the right of the upper two figure parts.}
\end{figure}

We next examine the diode rectified response, \dIddIQCL, of both samples and compare this to the laser emission illustrated in Fig. 4. This is plotted (solid lines) in Figs. 4(a) and (b) for Samples A and B, respectively. Also shown in Figs. 4(a) and (b) is the slope of the total emission intensity as a function of QCL current, $\partial P_{tot}/\partial I_{QCL}$ (dashed lines). All curves are normalized to unity to simplify comparison.

The total emitted power at fixed QCL bias $I_{QCL}$ is defined as
\begin{equation}
P_{tot}=\sum_m P_m
\end{equation}
where $P_m$ is the power emitted at the frequency of FP mode $m$. If a constant laser-diode coupling coefficient $C_m$ is assumed for each mode, and the diode rectified response is linear in power, then the diode response can be related to the power of individual modes by
\begin{equation}
\delta I_{D}/ \delta I_{QCL} = \sum_m C_m \: \partial P_{m}/\partial I_{QCL}.
\end{equation}
Considering equations (3) and (4), if $C_m$ are identical for all modes then the summation can be pulled through the differential so that $\delta I_{D}/ \delta I_{QCL} = C\:\partial P_{tot}/\partial I_{QCL}$. For Sample A there is a significant discrepancy between \dIddIQCL and $\partial P_{tot}/\partial I_{QCL}$. We conclude that the coupling strengths of individual FP modes to the diode are widely varied in this device. For Sample B, \dIddIQCL more closely tracks $\partial P_{tot}/\partial I_{QCL}$, although the discrepancies indicate there is still some asymmetry in the coupling strengths $C_m$ from mode to mode in this device.

\begin{figure}
\includegraphics[scale=0.95]{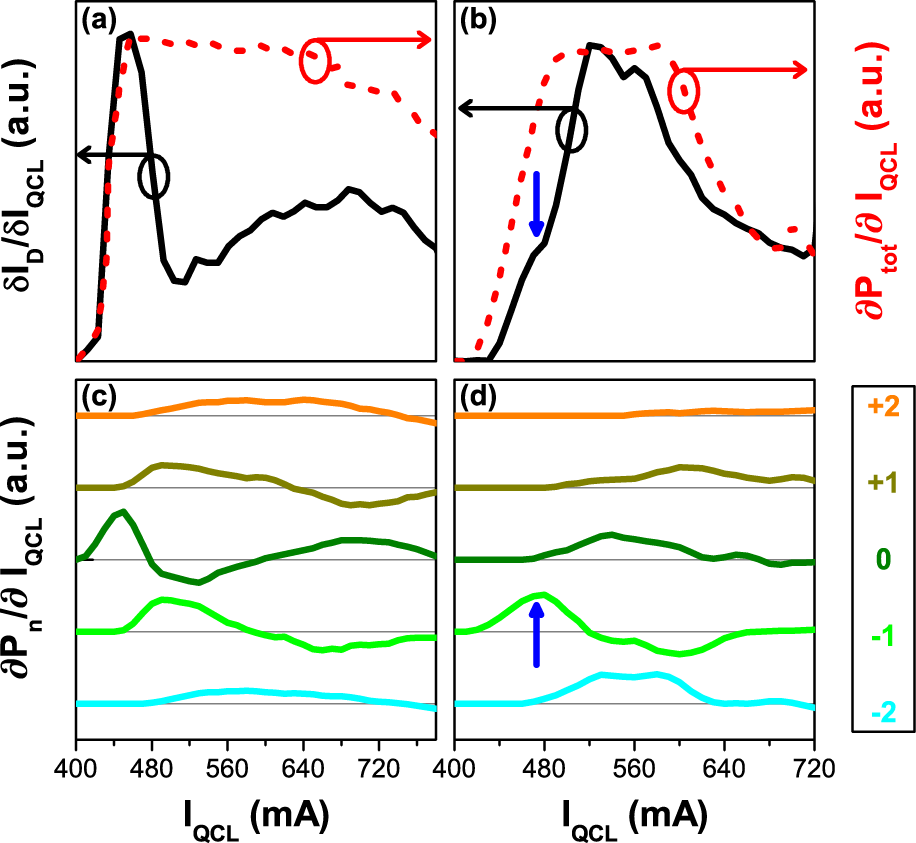}
\caption{The rectified diode response \dIddIQCL of Samples A and B is shown in (a) and (b), respectively, as a solid black line.  The slope of the total emitted power $\partial P_{tot}/\partial I_{QCL}$ is also shown as a red dashed curve.  Both sets of curves are normalized to unity.  The slope of emission intensity $\partial P_{m}/\partial I_{QCL}$ for modes 0, $\pm 1$ and $\pm 2$ is shown for Samples A and B in (c) and (d). The color coding to indicate the frequency of each mode $\pm$m correlates with the key in Fig. 3.}
\end{figure}

To better qualitatively understand the relative coupling strengths of the FP modes, the slopes of the power in modes 0, $\pm 1$ and $\pm 2$ as a function of QCL current, $\partial P_{n}/\partial I_{QCL}$, are shown in Figs. 4(c) and (d) for Samples A and B, respectively.  These numerical derivatives were calculated from the modal emission spectra in Figs. 3(c) and (d). It is useful first to examine the rectified response when only a single lasing mode is active. The diode response \dIddIQCL of Sample A peaks near $I_{QCL}=450$ mA where a maximum in the slope of its threshold mode, $\partial P_{0}/\partial I_{QCL}$, is also evident. For Sample B, however, the maximum in the slope of its threshold mode, $\partial P_{-1}/\partial I_{QCL}$, corresponds to a shoulder in \dIddIQCL at $I_{QCL}=475$ mA, as highlighted by the arrows in Figs. 4(b) and (d). The relative coupling strengths of the respective threshold mode fields to the diode evidently are very different in this pair of transceivers.

Next, we consider the rectified diode response when several FP modes couple simultaneously. In Sample A, a minimum in \dIddIQCL occurs at $I_{QCL}=500$ mA where the slope of mode 0's instantaneous power, $\partial P_{0}/\partial I_{QCL}$, is negative. Here the positive contributions from modes $\pm 1$, $\partial P_{\pm 1}/\partial I_{QCL}$, offset the negative contribution of mode 0, so that the net diode rectified response is still positive. However, there is a large discrepancy between \dIddIQCL and $\partial P_{tot}/\partial I_{QCL}$ as shown in Fig. 4(a),  indicating that modes $\pm 1$ have a smaller coupling coefficient than mode 0 does to the diode.

In Fig. 4(b), a maximum in \dIddIQCL for Sample B is found at $I_{QCL}=520$ mA. This critical feature is correlated with the maxima of $\partial P_{0}/\partial I_{QCL}$ and $\partial P_{-2}/\partial I_{QCL}$ in Fig. 4(d). The contributions of modes 0 and -2 produce the maximum rectified diode response from Sample B, rather than the fields of mode -1, the threshold mode, at its peak. In this device, the onset of multimode lasing does not correlate with weaker coupling of the laser fields to the diode.

Eq. 4 suggests that the coupling coefficients can be extracted from a least squares fit of the data. However, assuming constant $C_m \geq 0$, Eq. 4 could not be fit to the lasing modes over the entire range of QCL biases measured for either transceiver. This suggests that, not only are $C_m$ not equal, but also $C_m$ are not independent of $I_{QCL}$. This is consistent with prior work where \dIddIQCL decreased precipitously despite a relatively monotonic increase in the single mode laser emission intensity\cite{Dyer2013OPEX}. The diode response is generally not linear in the total or an individual mode's laser power over the measured QCL biases. Because the threshold mode of Sample A is relatively strongly coupled to the diode, it is possible that the diode is overpumped when the emission intensity increases. Changes in the mode shape, such as by Kerr lensing in the active region of the laser as seen in other lasers\cite{Mork1994,Paiella2000}, may also affect the coupling by changing confinement of the fields in the laser core.

In this letter, we have studied the rectified response of a Schottky diode embedded in a THz QCL. There is evidence that the diode couples non-identically to the FP modes in the QCL cavity as a byproduct of its subwavelength scale and relative location in the QCL waveguide. While a single mode laser is better for most heterodyne applications, in the case of a multi-mode laser, the diode should be placed where the modes are spatially in-phase to optimize the overlap of all FP mode field distributions with the diode. This occurs near the laser facets where the boundary conditions bring all modes spatially into phase. For heterodyne mixing, this would ensure that regardless of the specific FP mode that functions as the LO, there would be significant overlap of the LO fields with the diode. However, the integrated diode is in general not a linear power detector, even when accounting for the mode-to-mode differences in coupling strength. The reason for the non-constant coupling strength is not yet understood, but might be a result of changes in mode shapes in the cavity or overdriving of the diode. Further understanding is needed to optimize the overall responsivity, both rectified and IF, of the integrated diode for spectroscopic, imaging and communication applications.

%
%
%
\begin{acknowledgments}
This work was supported by the Sandia laboratory directed research and development (LDRD)
program. Sandia National Laboratories is a multi-program laboratory managed and operated by Sandia Corporation, a wholly owned subsidiary of Lockheed Martin Corporation, for the U.S. Department of Energy’s National Nuclear Security Administration under contract DE-AC04-94AL85000. 
\end{acknowledgments}
%

\end{document}